\documentclass[12pt,a4paper]{article}

\usepackage{epsf,graphicx,amssymb}

\newcommand{\be}{\begin{equation}}  
\newcommand{\ee}{\end{equation}}  
\newcommand{\ba}{\begin{eqnarray}}  
\newcommand{\ea}{\end{eqnarray}}  
\newcommand{\bi}{\begin{itemize}}  
\newcommand{\ei}{\end{itemize}}  
\newcommand{\bc}{\begin{center}}  
\newcommand{\ec}{\end{center}}  

\begin{document}
  
\title{On the stationary state of a network of inhibitory spiking neurons}

\author{Wolfgang Kinzel \thanks{ email: kinzel@physik.uni-wuerzburg.de}
\\ University of W\"urzburg \\  Institute of
  Theoretical Physics and Astrophysics\\ Am Hubland, W\"urzburg, Germany}

\maketitle

\abstract{ The background activity of a cortical neural network is
  modeled by a homogeneous integrate-and-fire network with unreliable
  inhibitory synapses. Numerical and analytical calculations show that
  the network relaxes into a stationary state of high attention. The
  majority of the neurons has a membrane potential just below the
  threshold; as a consequence the network can react immediately - on
  the time scale of synaptic transmission- on external pulses. The
  neurons fire with a low rate and with a broad distribution of
  interspike intervals. Firing events of the total network are
  correlated over short time periods. The firing rate increases
  linearly with external stimuli. In the limit of infinitely large
  networks, the synaptic noise decreases to zero. Nevertheless, the
  distribution of interspike intervals remains broad.}

\section{Introduction}

A neural network processes information by cooperating neurons, which
interact by exchanging spikes of action potentials. Numerous
experimental and theoretical findings suggest that a cortical neural
network has a background activity of neurons firing irregularly at low
rates \cite{abeles,gerstner}. On top of this noisy activity, information is
processed. It is still not known how a neural network codes, stores
and processes information. For instance, information in the cortex may
be coded in patterns of neuronal firing rates (Hebbian cell
assemblies) or in spatiotemporal patterns of spikes (synfire chains).
But, evidently,  all this occurs in a state of noisy activity.
Therefore it is important to understand the properties of the
background activity.

Quantitative estimates of the rate and size of the synaptic events in
the cortex suggest that the irregular background activity stems from a
balance between excitatory and inhibitory postsynaptic currents
\cite{shadlen,vreeswijk}. Both of these currents are large, but they
compensate and the activity of the network is triggered by the
fluctuations of the total current. Recent model calculations showed
how to integrate Hebbian cell assemblies or synfire chains in a
balanced network \cite{aviel,aviel2}.

However, experimental evidence for balanced synaptic input is still
not available and it is not understood whether and how a neural
network is able to compensate large synaptic currents of different
directions \cite{shadlen,trevelyan}. In addition, inhibitory pulses
can shunt excitation very effectively. Therefore I follow a different
approach and investigate a network with inhibitory synapses, only.  At
a first step, the investigations concentrate on one single mechanism:
unreliable inhibition. 

It is text book knowledge that synaptic transmission is a stochastic
process. In fact, experiments on single synapses indicate that
synapses transmit siganls with a probability which can be as low as a
few percent \cite{abeles,unreliable}. Only if an incoming spike is
repeated in a short time interval, the transmission probability
increases \cite{lisman}. This indicates that  the background activity is 
driven by unreliable synapses whereas for information processing teh
network improves synaptic reliability.    

In this paper I investigate how unreliable inhibition
effects the collective properties of the network, hence I do not
consider any kind of additional spatial or temporal noise, neither
disorder in the synaptic connections and thresholds nor noise in the
external stimulus.

One of the simplest models to describe the cooperative properties of
interacting neurons is a network of integrate-and-fire (IF) units
\cite{gerstner}. A single neuron is 
driven by an excitatory input and generates and transmits 
spikes. The total network may have unexpected complex properties which
are not obvious from a single model neuron.

From a more general point of view, IF networks are networks of pulse
coupled oscillators, and a large amount of work has been devoted to
the investigation of the dynamics of nonlinear oscillators
\cite{strogatz}.  Several interesting phenomena have been found for
IF-networks, for instance synchronization \cite{mirollo}, phase
locking, clustering \cite{ernst}, fast global oscillations
\cite{brunel-hakim}, stochastic resonance, deterministic and
transient chaos \cite{timme,zillmer} and 
  
In this paper I investigate the stationary state of a network of
inhibitory neurons. The effect of unreliable inhibitory synapses on
the global properties of the network is studied numerically as well as
analytically.  A single synapse consists of stochastic components; it
transmits signals with a probability which can be as low as a few
percent \cite{abeles,unreliable,lisman}.  Any other microscopic mechanisms
are omitted in the model, like synaptic delay, distribution of
thresholds and synaptic strengths. The effect of excitatory input is
modeled by a constant stimulus.

Neither the model nor the analytic methods of this paper are new.
Nevertheless, a new phenomenon is observed which, to my knowledge, was
not discussed before: A network with inhibitory couplings relaxes to a
stationary state of high attention.  After a few spikes per neuron,
the distribution of membrane potentials is sharply peaked just below
threshold.  As a consequence, the system is able to react immediately,
i.e. on the time scale of spike generation, to tiny external pulses.

In the model of this paper the synaptic noise disappears in the limit
of infinitely large networks. However, the system adjusts to a
stationary state where the distribution of interspike intervals
remains broad.

\section{The model}

In our model, each neuron is described by the following differential
equation for the membrane potential $V(t)$ \cite{gerstner},
\be
\label{e1}
\tau \frac{dV}{dt}=-V(t)+V_\infty 
\ee 
$\tau$ is the relaxation time
for the membrane potential and the driving potential $V_\infty$ models
a constant excitatory stimulus. When the membrane potential $V(t)$
reaches the threshold value $\theta$ the neuron fires a spike and
resets its potential to the value $V_r$.

A homogenous network is studied where each neuron
has the parameters $\tau=10$ ms, $V_r=-70$ mV, $V_\infty=-50$ mV and
$\theta=-51$ mV. Hence, without any interactions, each neuron would
fire periodically with the time period $T= \tau \ln 20 \simeq 30$ ms
which corresponds to a frequency of 33 Hz.

I consider a homogeneous network of $N$ mutually coupled neurons. If any
neuron fires, it sends its spike to its synapses which are connected
to all other neurons. The synapse is unreliable, it transmits the
signal with a probability $p$, only. If a signal is transmitted, it
reduces the postsynaptic potential by an amount $J$. In our
simulations, I use the parameters $N=10000$, $p=0.5$ and $J=0.002$ mV.
Hence, each synapse is extremely weak; but the effective strength $p J
N=10$mV  is of the order of the  difference between threshold
and reset potentials.

Between two consecutive pulses in the total the network, each neuron follows
Eq. (\ref{e1}) which is easily solved.  The time needed to
increase the potential $V(t_1)$ to a value $V(t_2)$ is given by 
\be
\label{e2}
t_2-t_1= \tau \, \ln \frac{V_\infty - V(t_1)}{V_\infty - V(t_2)} 
\ee

This equation allows a simple numerical simulation of the total
network without solving any differential equation. At each
computational step, the algorithm searches for the neuron with the maximal
values of $V(t)$, calculates the time this neuron needs to reach the
threshold $\theta$ and resets its potential to the value $V_r$. Then
the potentials of all other neurons are calculated from Eq.(\ref{e2})
and, with probability $p$, their values are reduced by the amount of
$J$. This process is iterated until the system has reached a
stationary state. Then the distributions of potentials and interspike
intervals are recorded.

\section{Results of the  simulations}

The homogeneous network of IF-neurons with the parameters given before
has been simulated numerically. I start from a flat distribution of
potentials between reset and threshold value. After a few spikes per
neuron the network has lost the memory to its initial state and has
relaxed to a stationary state. Note that our model is stochastic,
hence I have to discuss \emph{distributions} of quantities of interest.

\begin{figure}[h]
\begin{center}
\includegraphics[width=.6\textwidth]{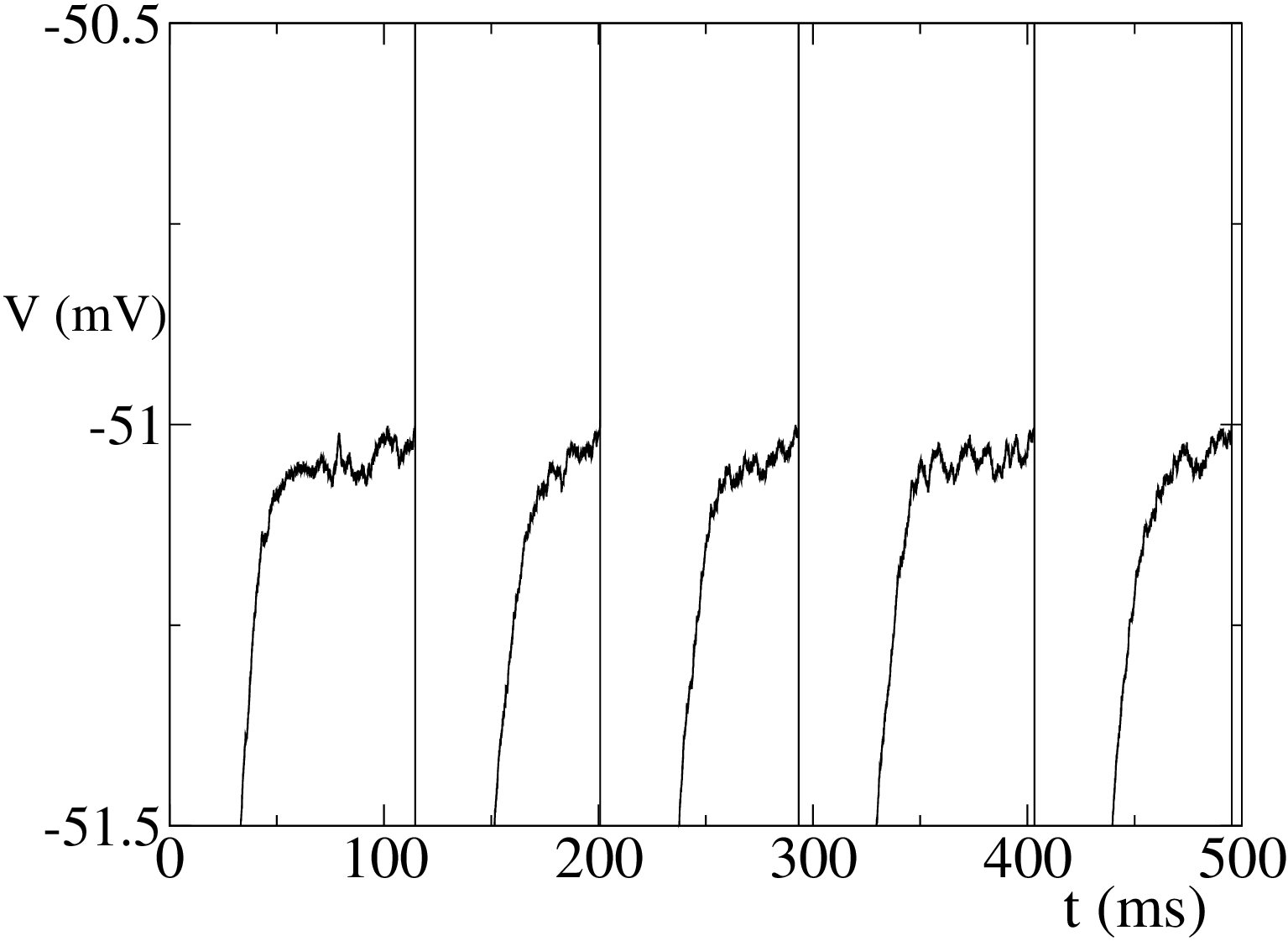}
\end{center}
\caption{Membrane Potential $V(t)$ as a function of time for a network
  of $10000$ inhibitory neurons with unreliable synapses. As soon as
  the potential reaches the threshold value $\theta=-51$mV the neuron
  emits a spike and resets its potential to the value
  $V_r=-70$mV. Between two spikes,the potential is reduced about 5000
  times by the value of $J=0.002$mV }
\label{ut}
\end{figure}

In the following I show results for the distribution of membrane
potentials and interspike intervals, both of which are accessible for real
neurons. Fig.\ref{ut} shows the membrane potential of one single
neuron as a function of time. Since the network is homogeneous all
neurons have identical properties. Between two spikes, each neuron
receives about  5000 pulses from the 9999 other neurons of the
network. Each pulse reduces its potential by the amount of $J=0.002$ mV. 

\begin{figure}[h]
\begin{center}
\includegraphics[width=.6\textwidth]{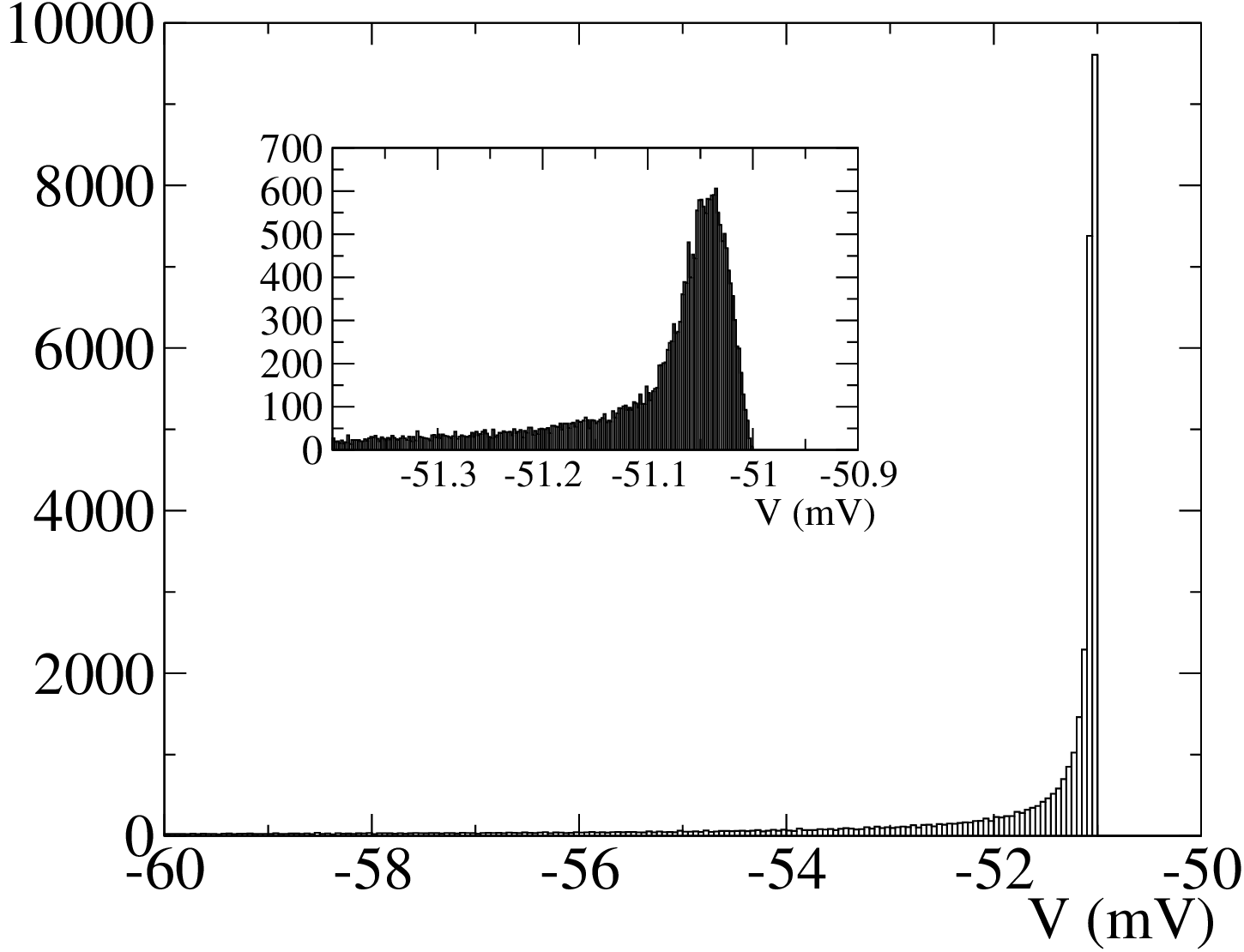}
\end{center}
\caption{ Histogram of the membrane potentials of the network at some
  specified time. Initially, the distribution was flat between -70mV
  and -51 mV. In the stationary state, shown here, most of the neurons
have a potential immediately below threshold threshold. The inset
shows the same distribution close to threshold.}
\label{u-hist}
\end{figure}
  
I see that a typical neuron increases its potential exponentially
fast to a kind of plateau immediately below threshold. The stochastic
input leads to fluctuations of the membrane potential. Whenever this
potential exceeds the threshold value the neuron fires. Note that the
average spike interval increases from 30 ms without couplings to about
100 ms. Obviously, the inhibitory pulses reduce the firing rate. But
less obvious, most of the neurons have a membrane potential close to
the threshold value. This is shown in Fig. \ref{u-hist}. The distribution
of membrane potentials has a peak which is only 0.05 mV below threshold. 

\begin{figure}[h]
\begin{center}
\includegraphics[width=.6\textwidth]{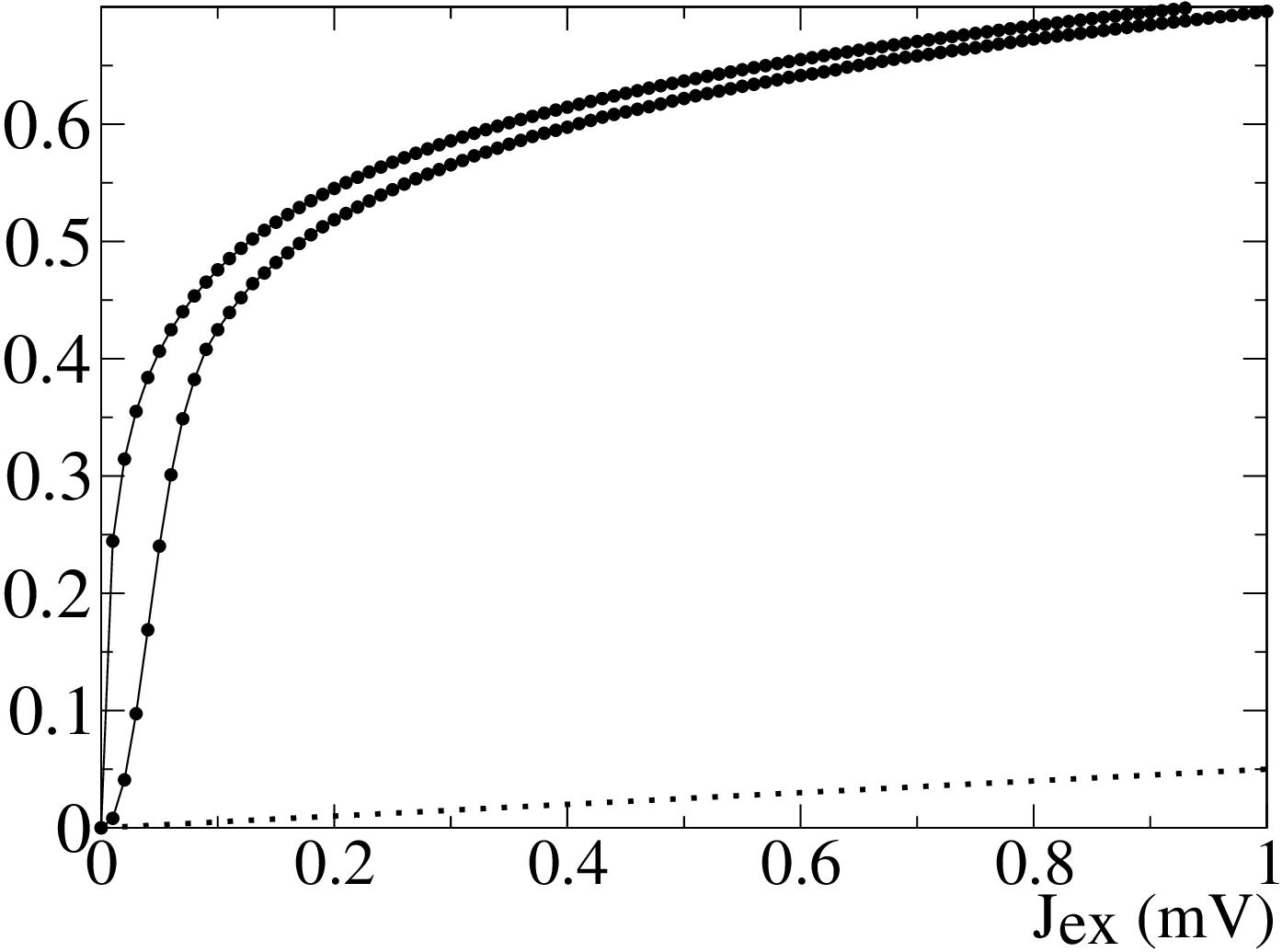}
\end{center}
\caption{The fraction of neurons which fires immediately after an excitatory pulse
  of a given value $J_{ex}$. The network with stochastic synapses (lower values) is
  compared with a corresponding deterministic network (upper
  values). The dotted line shows the corresponding fraction of neurons
  for the initial state with a flat distribution of potentials.}
\label{sens}
\end{figure}

\begin{figure}[h]
\begin{center}
\includegraphics[width=.6\textwidth]{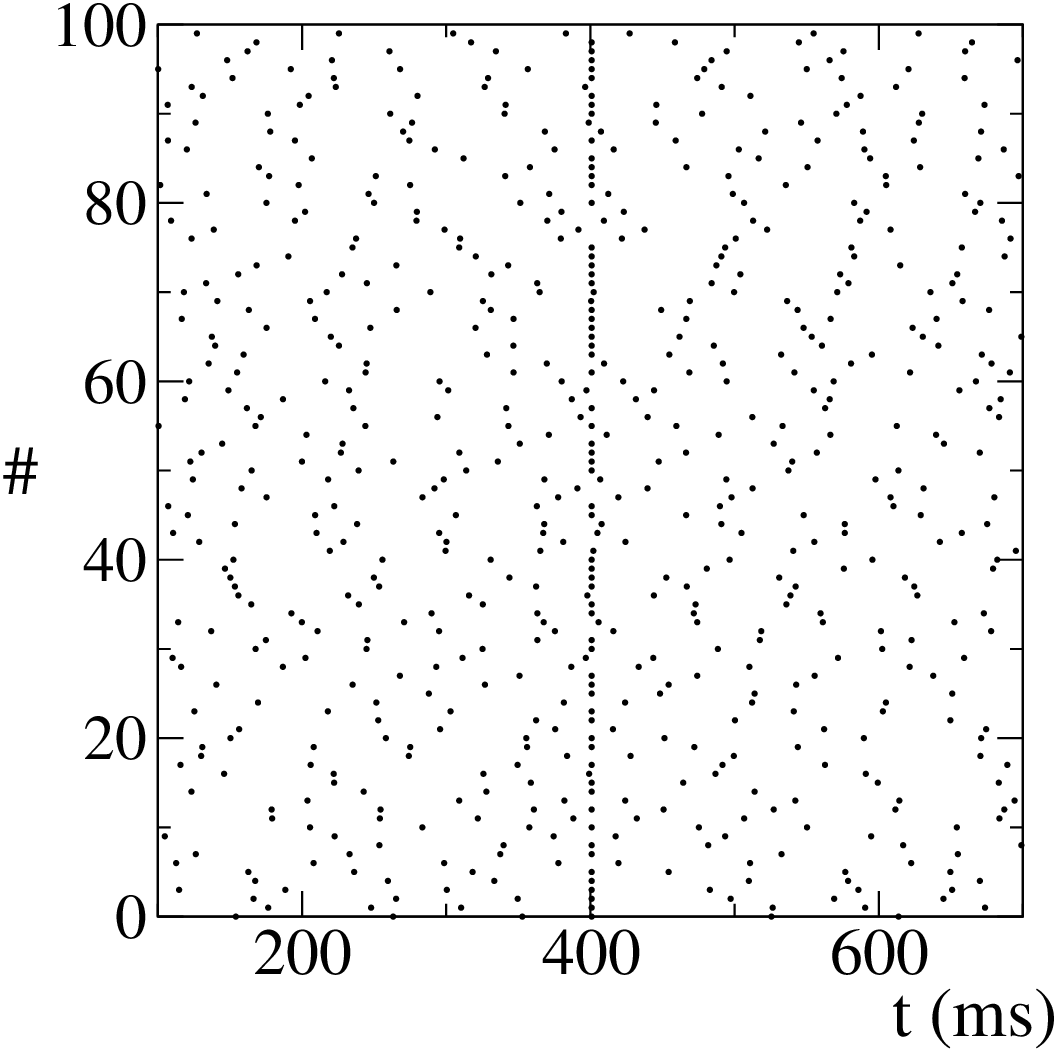}
\end{center}
\caption{Spike patterns of one hundred neurons (out of 10000). At
  time $t=400$ms, an excitatory pulse is applied which increases all
  postsynaptic potentials by the amount of 0.5 mV. According to the
  previous figure \ref{sens}, about 60\% of the neurons fire immediately.}
\label{pattern}
\end{figure}

\begin{figure}[h]
\begin{center}
\includegraphics[width=.6\textwidth]{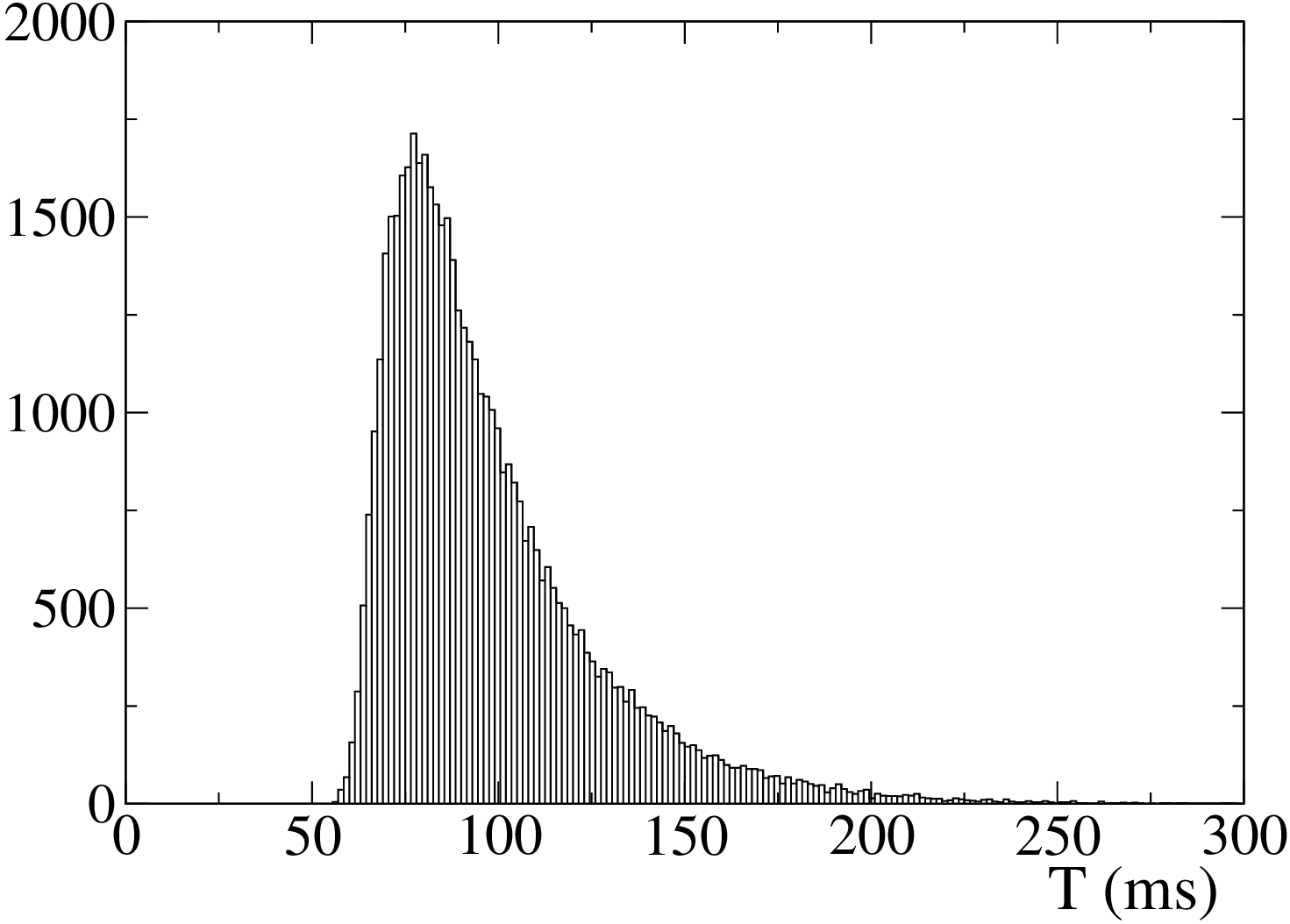}
\end{center}
\caption{ Histogram of the intervals between spikes of a single neuron.}
\label{sp-int}
\end{figure}

Note that the initial distribution was flat between -70 mV and -50 mV.
After a short time the system has relaxed to a stationary where most
of the neurons have a potential immediately below threshold. As a
consequence, a large fraction of neurons can react immediately to tiny
pulses of the order of a fraction of millivolts. In Fig \ref{sens} the
fraction of neurons is shown which fires immediately after a
excitatory pulse of a given strength $J_{ex}$. For comparison, I have
included the corresponding curve for the initial state. Fig.
\ref{pattern} shows the corresponding spike pattern for
$J_{ex}=0.5$mV. All of the neurons which have responded to the pulse
excitation remain quiet for at least 50 ms, see Fig \ref{sp-int}.
These results show that the network has relaxed to a state of high
attention and can react to incoming small pulses on the time scale of
synaptic transmission.

\begin{figure}[h]
\begin{center}
\includegraphics[width=.6\textwidth]{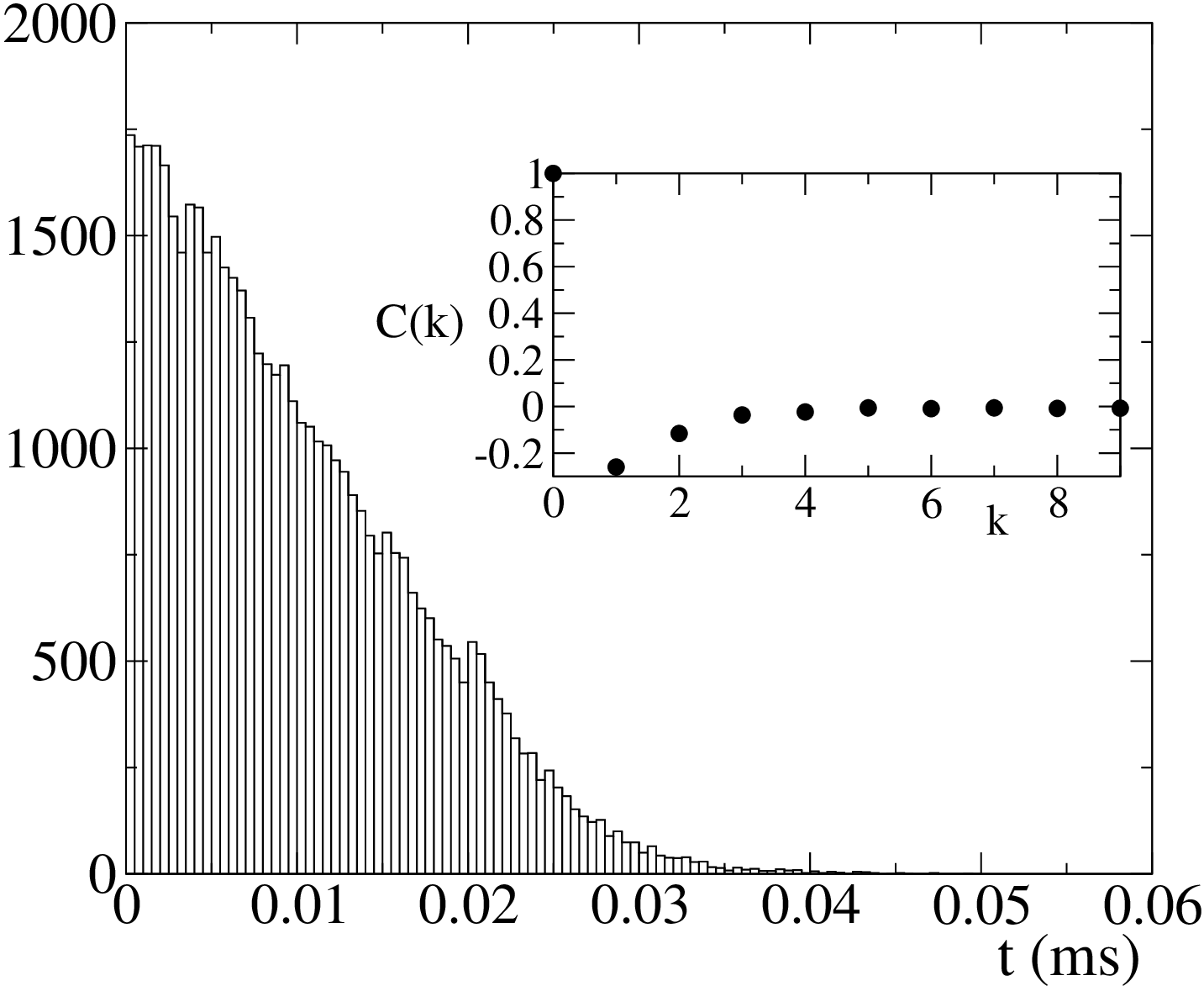}
\end{center}
\caption{ Histogram of the intervals between spikes of the total
  network. The inset shows the corresponding cross correlation of consecutive
spikes. }
\label{ev-int}
\end{figure}

The distribution of interspike intervals $T$ is shown in Fig. \ref{sp-int}.
There are no intervals below 50 ms and the distribution decays
exponentially fast for large values of $T$. Its mean value is
$<T>=97.6 $ ms and its standard deviation $\sigma= 29.0$ms, which
gives a CV-value of 0.3.  Since the network has identical neurons, the
average time  $s$ between two consecutive spikes in the total network is
$<s> = <T>/N \simeq 0.01$ ms.  Usually one assumes that the spikes
arriving at a single neuron are uncorrelated, which would yield a
Poisson, i.e. an exponential distribution of the intervals $s$ between
synaptic events.  Fig. \ref{ev-int} shows that this is not true.
Although small values of $s$ are more frequent than larger ones, the
distribution is not Poissonian indicating correlations between events.
In fact, the inset of Fig. \ref{ev-int} shows the correlations between
consecutive  spikes. The spike intervals of the
total network are ordered, $(s_1,s_2,s_3,...s_j,..)$, and
Fig. \ref{ev-int} shows the cross correlation
\be
C(k)=\frac{<s_{j+k}s_j> - <s_j>^2}{<s_j^2>-<s_j>^2} 
\ee  

Correlations can also be observed from the statistics of spike
counts. For instance, the Fano factor $F$ is defined as the ratio of
the variance divided by the average number of spikes observed in some
time interval $\Delta T$. For a Poisson process one has $F=1$
independent of $\Delta T$. I have recorded the number of network events 
for $\Delta T=1$ms and find a much smaller Fano factor of $F=0.15$. 
Hence the distribution of the number of spikes is much smaller than the
corresponding one for uncorrelated synaptic inputs.

\begin{figure}[ht]
\begin{center}
\includegraphics[width=.6\textwidth]{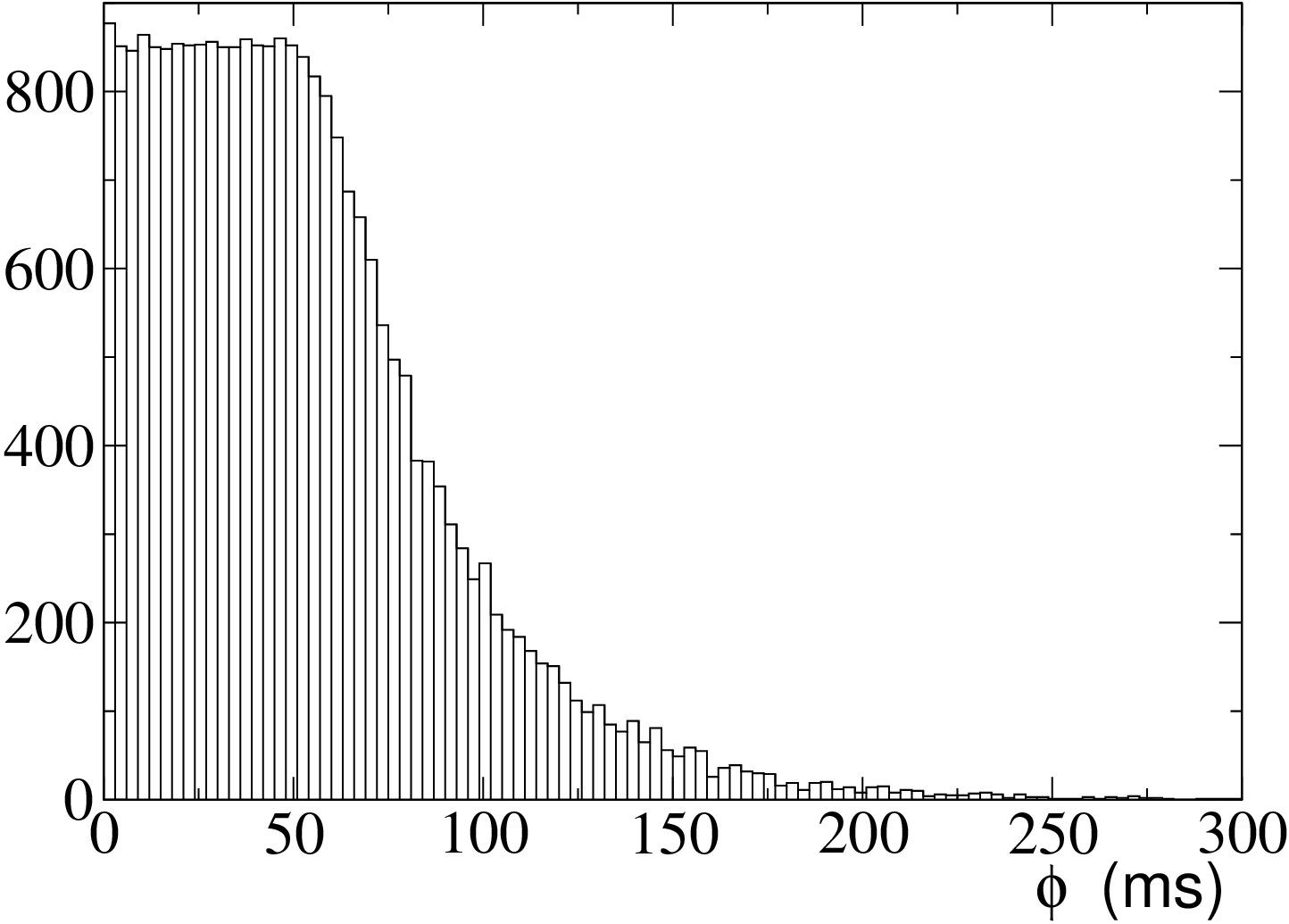}
\end{center}
\caption{Histogram of the phases of the neurons.}
\label{phase}
\end{figure}

In the context of pulse coupled oscillators the phase $\phi$ of an
oscillator is of interest. When a reference neuron fires at time $t$
the phase is defined as $\phi_k=t-t_k$ where neuron $k$ has fired at
at previous time $t_k$. The distribution of phases is shown in Fig.
\ref{phase}. The distribution of a deterministic network which is flat
between 0 and 100 ms (see below) is rounded due to synaptic noise.

It may be interesting to compare the properties of our stochastic
model with the corresponding deterministic one, i.e. with the model
with parameters $p=1$ and $J=0.5$.  The effective strength $p J N =10$
mV is identical but the synapses function completely reliable.  Our
simulations show that the deterministic model relaxes to a periodic
state with a flat distribution of phases.  Each neuron fires
periodically with the period $T=100.04$ ms. The time $s$ between
events of the total network is constant, $s=T/N=0.10004$ ms. Hence,
both of the distributions of interspike as well as interevent
intervals have a sharp peak at a single value. This property of the
deterministic network is definitely at variance with the behavior of
real neural networks.

Nevertheless, the distribution of membrane potentials is similar to
the one of the corresponding stochastic network, Fig. \ref{u-hist}. Only
very close to threshold one observes deviations. Note that a flat distribution
of phases transforms to a distribution of potentials which is
proportional to $1/(\mu-u)$ with a cutoff above the value
$\theta$. The parameter $\mu$ is slightly larger than the value
$\theta$ and  will be calculated in the following
section.

\section{Analytic approximations}

Although our model is rather simple it cannot be calculated
analytically. Even two neurons with unreliable synapses yield a
complex multifractal distribution of interspike intervals which can
only be calculated numerically \cite{kestler}. But analytic
approximations are possible and help to understand the influence of
the model parameters on the behavior of the network.

It is well known that a stochastic network may be approximated by a
diffusion process of the membrane potential \cite{tuckwell,gerstner,brunel}.
Assuming uncorrelated synaptic input one can describe the time
dependent distribution of potentials by a Fokker-Planck equation for
an Ornstein-Uhlenbeck process.  Spike intervals correspond to first
passage times.  Although, even for this approximation, there is no
closed expression for the distribution of spike intervals, one can
calculate its mean value from the solution of an ordinary differential
equation.

Two parameters enter this approximation: The mean value and the
variance of the synaptic input. Let us first ignore the fluctuations
and consider the membrane potential, Eq.(\ref{e1}), averaged over the
stochastic synaptic input.  Each spike in the network adds a pulse $p J$ to the
potential of every neuron. The average time between two pulses is given by
$<s>=<T>/N$, hence I obtain the following equation
\be
\tau \frac{dV}{dt}=- V(t) +V_\infty-  \frac{\tau p J N}{<T>} 
\ee
The mean value of the membrane potential moves exponentially fast  to the value
\be
\label{e5}
\mu=V_\infty- \frac{\tau J g N}{<T>}
\ee
until it crosses the threshold.
If I would neglect fluctuations, the spike interval would follow from
Eq.(\ref{e2}), which yields
\be
\label{e6}
\frac{<T>}{\tau}= \ln \frac{V_\infty-  \frac{\tau p J N}{<T>}-V_r}{V_\infty- 
  \frac{\tau p J N}{<T>}-\theta} 
\ee
This self-consistent equation for the interspike interval is exact for the
deterministic model in the limit of a large network, $N\to \infty$
with $J N=$const. Note that in this case, the system relaxes to a
periodic state with   only one single
spike interval $T$. If the effective  synaptic strength $JN$ 
is large,
i.e. if the spike interval $T$ is much large than $\tau$, the
denominator of Eq.(\ref{e6}) approaches zero and one obtains 
\be
\label{e7}
\frac{<T>}{\tau}=\frac{p J N}{V_\infty-\theta}
\ee
For a given value of $J$, the spike interval is proportional to the
size of the network and its inverse, the spike rate, is proportional
to the difference between stimulus and threshold potential.

Our numerical simulations showed that the deterministic network relaxes
to a state with a flat distribution of phases (times after
firing). Using this fact, one immediately derives the distribution of
membrane potentials from Eq.(\ref{e1}) with the result
\be
\label{e8}
p(V)= \frac{\tau N}{T} \frac{1}{\mu-V} \quad \mbox{for}\quad
V_r<V<\theta \ee 
In the limit of infinitely large networks the
potential $\mu$ approaches the threshold, hence in this limit almost
all neurons have a potential very close to threshold. In fact,
Eqs.(\ref{e6},\ref{e7}) give the fraction $P(\varepsilon)$ of neurons
which have a membrane potential in the interval
$[\theta-\varepsilon,\theta]$, 
\be
P(\varepsilon)=\frac{\ln(\mu-\theta+\varepsilon)-\ln(\mu-\theta)}
{\ln(\mu-V_r)-\ln(\mu-\theta)}
\ee
This fraction approaches $P(\varepsilon)=1$ with $\mu \to \theta$.

For the stochastic version of the model, the simulations of the
previous section yielded a broad distribution of spike intervals $T$
which is caused by the fluctuations of synaptic inputs. Therefore, in
the analytic approximation, fluctuations of the synaptic input are
included by adding noise to Eq.(\ref{e1}) with mean value $\mu$ and
variance $\sigma^2$. The value of the mean $\mu$ is given by
Eq.(\ref{e5}) and the variance can be estimated by assuming a Poisson
statistics, one finds \cite{brunel} 

\be \mu=V_\infty- \frac{\tau J g
  N}{<T>},\quad\sigma^2= J^2 \frac{p N \tau}{<T>} \ee The mean spike
interval is obtained from the self-consistent equation \be
\label{e10}
\frac{<T>}{\tau}=\int\limits_0^\infty du \, {\rm e}^{-u^2}
\left[\frac{{\rm e}^{2 y_\theta u} -\mbox{e}^{2 y_r u}}{u}\right] 
\ee
with
\be
y_\theta=\frac{\theta-\mu}{\sigma},\quad y_r=\frac{V_r-\mu}{\sigma}
\ee
If $y_\theta$ is large, i.e. if the fluctuations of the potential are
small compared to the difference between the driving potential $\mu$
and the threshold $\theta$, then the mean first passage time can be
approximated by 
\be
\label{e12}
\frac{<T>}{\tau} \simeq \rm{e}^{y_\theta^2}
\ee  
For the parameters of our simulations of the previous section,
Eqs. (\ref{e10},\ref{e12}) give $<T>=94.6$ ms and $<T>=93.5$ ms which is
in reasonable agreement with the mean spike interval  of the
simulations, $<T> = 97.6$ ms. The effective average  stimulus $\mu$ is
only 0.05 mV below the threshold potential $\theta$, but this
difference is still large compared to the fluctuations, one finds 
$y_\theta=1.23$.

Note that, contrary to the deterministic case, the mean potential
relaxes to a value $\mu$ \emph{below} the threshold $\theta$. The spikes are
generated by fluctuations which cross the threshold. Nevertheless, in
both cases,   the value of $\mu$ is close to  $\theta$ 
which almost gives an identical average spike interval, Eq.(\ref{e6}).
In particular, the spike rate increases linearly with the stimulus $V_\infty$.
 
It might be interesting to consider the limit of large networks, $N
\to \infty$ with constant $J N=C$. Inserting the average spike interval, Eq.(\ref{e6}),
into Eq.(\ref{e12}) yields
\ba
\left(\frac{\theta-\mu}{\sigma}\right)^2 & = & \ln
\frac{pC}{V_\infty-\theta} \\
\sigma^2 &=& \frac{C }{N} (V_\infty-\theta) \\
(\theta-\mu)^2 &=& \frac{C}{N}\,(V_\infty-\theta)\,\ln\frac{pC}{V_\infty-\theta}
\ea
Hence the fluctuations of the synaptic noise show an unusual
behaviour in this limit: Its variance \emph{decreases} with increasing
system size $N$. However, the fluctuations relative to the difference
between driving potential $\mu$ and threshold $\theta$ remain
constant, yielding a constant mean spike interval $<T>$.
The fluctuations do not disappear in the thermodynamic limit
$N \to \infty$. 

\section{Discussion}

A homogeneous neural network with purely inhibitory synapses relaxes
to a state of high attention. A large fraction of neurons have a
membrane potential which is just below threshold.  Hence the network
can immediately - on the time scale of synaptic transmission - react
on external excitatory pulses. 

Although, at any time, the majority of the neurons accumulate closely
below threshold, each single neuron fires irregularly with a low rate.
In our model the broad distribution of spike intervals stems from
synaptic unreliability; a synapse transmits the incoming spike with
some probability, only. The numerical simulations show that the spiking
events of the total network are correlated in time. Uncorrelated
synaptic noise still leads to correlated spike intervals, even for
infinite-range models.

For our model, in the thermodynamic limit any neuron receives an
infinite number of synaptic inputs. Usually, one argues that, in this
limit, the fluctuations of the synaptic input can be ignored. Our
model, however, shows that even for an infinite number of inhibitory
inputs the fluctuations of the membrane potential have a large effect.
The network adjusts itself such that there
remains a broad distribution of spike intervals in this limit.

The stationary state of the network is a pool of neurons firing
irregularly with a low rate. The firing rate increases linearly with a
global external stimulus.  The network can immediately react to
external pulses.  Our model contains unreliable inhibitory synapses,
only. Hence a network does not need to balance large excitatory and
inhibitory synaptic currents to achieve these properties.

Our investigations concentrated on a very simple model:
Integrate-and-fire neurons with unreliable inhibitory synapses which
couple all neurons. Each neuron has identical parameters.  One may
wonder whether and how the properties of the network change when the
model is extended to other mechanisms. In fact, much is known about
IF-networks. For instance, synaptic delay leads to clusters of
synchronized activity \cite{ernst} and to oscillating firing rates
\cite{brunel-hakim}. A distribution of the number of short-range
synaptic contacts increases the correlations between consecutive spike
intervals \cite{hertz}. Excitatory synapses can produce synfire
chains, i.e. waves of synchronized activities moving through the
network \cite{abeles,aviel}. When the excitation is too strong, the neurons
synchronously fire at a  high rate \cite{brunel}. A distribution of threshold
values obviously leads to a distribution of firing rates. In fact, I
find that in this case, the stationary state of the  network contains a
fraction of neurons which never fire.

All these extensions modify the properties of state
of irregularly firing neurons. However, one of the main questions remains
unsolved: How does the network process information on top of this
background activity?

\noindent {\bf Acknowledgement}

The author would like to thank Moshe Abeles, Ramon Huerta and Ido
Kanter for useful comments.


\begin{thebibliography}{8.}

\bibitem{abeles} Abeles M (1991) Corticonics, Cambridge University Press

\bibitem{unreliable} Allen C  and Stevens CF (1994) An evaluation of
  causes for unreliability of synaptic transmission, Proc. Natl. Acad.
  Sci. USA 91: 10380-10383 

 
\bibitem{aviel} Aviel Y, Mehring,  Abeles M and Horn D (2003) 
On embedding synfire chains in a balanced network, Neural Comp. 15(6):
1321-1340 

\bibitem{aviel2} Aviel Y, Horn D, Abeles M (2005) 
Memory capacity of balanced networks, Neural Comp. 17: 691-713 

\bibitem{brunel} Brunel N (2000) Dynamics of sparsely connected networks
of excitatory and inhibitory spiking neurons, J. Comp. Neuroscience 8:
183-208

\bibitem{brunel-hakim} Brunel N and Hakim V (1999) Fast global oscillations
  in networks of integrate-and-fire neurons with low firing rates,
Neural Computation 11: 1621-1671 

\bibitem{ernst} Ernst U, Pawelzik K and Geisel T (1995) Synchronization
  induced by temporal delays in pulse-coupled oscillators,
  Phys. Rev. Lett. 74: 1570-1574 

\bibitem{gerstner} Gerstner W and Kistler W (2002) Spiking Neuron
  Models, Cambridge University Press 

\bibitem{hertz} Hertz JA, Richmond BJ and Nilsen K (2003) Anomalous
  response variability in a balanced cortical network, Neurocomp.
  52-54: 787-792 

\bibitem{mirollo}
Mirollo RE and Strogatz SH (1990) SIAM J. Appl. Math. 50: 1645 

\bibitem{kestler} Kestler J and Kinzel W (2006), Multifractal
  distribution of spike intervals for two neurons coupled by
  unreliable pulses,  J. Phys A 39: L461-L466
 
\bibitem{lisman} Lisman JE (1997) Bursts as a unit of neural
  information: making unreliable synapses reliable, Trends
  Neurosci. 20: 38-43

\bibitem{shadlen} Shadlen MN and Newsome WT (1994) Noise, neural codes
and cortical organization, Curr. Opin. Neurobiol. 4(4): 569-579

\bibitem{vreeswijk} Van Vreeswijk,C., and Sompolinsky,H., Chaos in
  neuronal networks with balanced excitatory and inhibitory activity,
  Science 274, 1724-1726 (1996)

\bibitem{strogatz} Strogatz SH (2001), Nonlinear dynamics and chaos,
  Cambridge University press

\bibitem{timme} Timme M, Wolf F and Geisel T (2002) Coexistence of
  regular and irregular dynamics in complex networks of pulse-coupled
  oscillators, Phys. Rev. Lett. 89: 258701 

\bibitem{trevelyan} Trevelyan AJ and Watkinson O (2005) Does inhibition
  balance excitation in neocortex? Progress in Biophysics and
  Molecular Biology 87 109-143


\bibitem{tuckwell} Tuckwell HC (1998) Introduction to theoretical
  neurobiology, Cambridge University Press 



\bibitem{zillmer} Zillmer R, Livi R, Politi A and Torcini A (2006)
  Desnychronized stable states in diluted neural networks, cond-mat/0608188 


\end{thebibliography}
\end{document}